\documentclass[sigplan,screen]{acmart}
\AtBeginDocument{%
  }

\usepackage{subcaption}
\usepackage{tikz}
\usepackage{algorithm}      % Required for algorithm environment
\usepackage[noend]{algpseudocode}
\usepackage{makecell}

\usetikzlibrary{automata, positioning, arrows}

\newcommand{\blue}[1]{{\color{blue}
{\textbf{#1}}}}

\setcopyright{none}
\acmConference[EGRAPHS'25]{}{June 2025}{Seoul, South Korea}
\begin{document}

%%
%% The "title" command has an optional parameter,
%% allowing the author to define a "short title" to be used in page headers.
\title{Cut Tracing with E-Graphs for Boolean FHE Circuit Synthesis}

%%
%% The "author" command and its associated commands are used to define
%% the authors and their affiliations.
%% Of note is the shared affiliation of the first two authors, and the
%% "authornote" and "authornotemark" commands
%% used to denote shared contribution to the research.
\author{Julien de Castelnau}
\email{julien.decastelnau@epfl.ch}
\orcid{0000-0001-9705-2007}
\affiliation{%
  \institution{EPFL}
  \city{Lausanne}
  \country{Switzerland}
}

\author{Mingfei Yu}
\email{mingfei.yu@epfl.ch}
\orcid{0009-0009-6816-8903}
\affiliation{%
  \institution{EPFL}
  \city{Lausanne}
  \country{Switzerland}
}

\author{Giovanni De Micheli}
\email{giovanni.demicheli@epfl.ch}
\orcid{0000-0002-7827-3215}
\affiliation{%
  \institution{EPFL}
  \city{Lausanne}
  \country{Switzerland}
}

%%
%% By default, the full list of authors will be used in the page
%% headers. Often, this list is too long, and will overlap
%% other information printed in the page headers. This command allows
%% the author to define a more concise list
%% of authors' names for this purpose.
\renewcommand{\shortauthors}{de Castelnau, Yu, De Micheli}

%%
%% The abstract is a short summary of the work to be presented in the
%% article.
\begin{abstract}

Fully Homomorphic Encryption (FHE) is a promising privacy-preserving technology enabling secure computation over encrypted data. A major limitation of current FHE schemes is their high runtime overhead. As a result, automatic optimization of circuits describing FHE computation has garnered significant attention in the logic synthesis community. Existing works primarily target the \textit{multiplicative depth} (MD) and \textit{multiplicative complexity} (MC) of FHE circuits, corresponding to the total number of multiplications and maximum number of multiplications in a path from primary input to output, respectively. In many FHE schemes, these metrics are the primary contributors to the homomorphic evaluation runtime of a circuit. However, oftentimes they are opposed: reducing either depth or complexity may result in an increase in the other. To our knowledge, existing works have yet to optimize FHE circuits for overall runtime, only considering one metric at a time and thus making significant tradeoffs. In this paper, we use e-graphs to augment existing flows that individually optimize MC and MD, in a technique called \textit{cut tracing}. We show how cut tracing can effectively combine two state-of-the-art MC and MD reduction flows and balance their weaknesses to minimize runtime. Our preliminary results demonstrate that cut tracing yields up to a 40\% improvement in homomorphic evaluation runtime when applied to these two flows.
\end{abstract}

%%
%% This command processes the author and affiliation and title
%% information and builds the first part of the formatted document.
\maketitle

\section{Introduction}

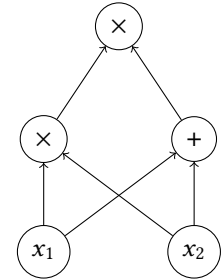
\begin{figure}
    \centering
    \begin{tikzpicture}
    \node[state, xshift=0cm, yshift=0cm, minimum size=0.5cm] (x1)   {$x_1$};
    \node[state, xshift=2cm, yshift=0cm, minimum size=0.5cm] (x2)   {$x_2$};
    \node[state, xshift=0cm, yshift=1.5cm, minimum size=0.5cm] (n3) {$\times$};
    \node[state, xshift=2cm, yshift=1.5cm, minimum size=0.5cm] (n4) {$+$};
    \node[state, xshift=1cm, yshift=3cm, minimum size=0.5cm] (n5) {$\times$};
    %\node (f) at (1,4.5) {$f$};
    
    \draw [->] (x1) edge (n3);
    \draw [->] (x2) edge (n3);
    \draw [->] (x2) edge (n4);
    \draw [->] (x1) edge (n4);
    \draw [->] (n3) edge (n5);
    \draw [->] (n4) edge (n5);
    %\draw [->] (n5) edge (f);
\end{tikzpicture}
    \caption{The expression $(x_1 \times x_2) \times (x_1 + x_2)$ as an FHE circuit. $x_1$ and $x_2$ are ciphertexts and $\times$, $+$ correspond to homomorphic multiply and add, respectively.}
    \label{fig:ckt}
\end{figure}

An FHE circuit is a DAG that describes computations on ciphertexts with the main operations of FHE, homomorphic add and multiply. Figure \ref{fig:ckt} shows an example circuit computing the expression $(x_1 \times x_2) \times (x_1 + x_2)$, where $x_1$ and $x_2$ are ciphertexts. The plaintext message encoded in the ciphertext is assumed to be an integer mod some $p$, termed the \textit{plaintext modulus}. When $p=2$, homomorphic add and multiply operations are equivalent to XOR and AND gates on Booleans. In this case, an FHE circuit is equivalent to an XOR-AND graph (XAG). We focus on the Boolean case in this work, because many FHE applications, such as a CRC check, rely on a bitwise integer encoding \cite{gouert2023sok}. 

The limiting factor in most FHE schemes is the runtime overhead of multiplication, with addition being comparatively cheap. Therefore, when considering an FHE circuit, the total number of multiplications, the MC, is a major performance indicator. However, multiplications also have a disparate impact on the \textit{noise budget} of the ciphertext, in comparison to addition. FHE schemes store a noise component in ciphertexts as part of a ring learning with errors (RLWE) construction, used to provide their security guarantees. Every operation on the ciphertext increases its noise; if the noise grows past a certain budget, decryption will fail. While it is possible to reset the noise on a ciphertext in a procedure known as \textit{bootstrapping}, it is often prohibitively time-consuming. Instead, the ciphertext size can be increased, raising the relative noise budget, at the cost of increasing the global overhead of all operations. From this, the constraint of multiplicative depth (MD) emerges: as each multiplication operation scales the noise quadratically, the longest path of multiplications in sequence in the circuit dictates the allocated noise budget, in turn increasing the global runtime cost. 

There is a rich body of work on automatic methods to reduce the MD and MC of Boolean circuits \cite{carpov2017multistart, lee2020lobster, haner2022esop, testa2020toolbox}. For example, \cite{lee2020lobster} uses program synthesis to learn a set of MD-optimal circuit rewrite rules, applying them towards a set of HE benchmarks to reduce their evaluation runtime. Likewise, \cite{testa2020toolbox} presents a flow for MC reduction using a combination of area-reducing logic synthesis techniques.  However, to our knowledge, the joint optimization of both MC and MD in the context of FHE has not been addressed. Existing techniques only consider one quantity at a time, often blowing up MC in the process of reducing MD, or vice versa.

These techniques are difficult to extend to a model of overall HE cost that balances MC and MD for two reasons. First, they all apply optimizations locally, relying on the computation of some local gain. When both MD and MC are incorporated, gain becomes a poor indicator of the global cost that will be achieved.
Second, they apply these optimizations as destructive rewrites to the circuit, meaning they compose poorly: alternating MD and MC optimization passes does not work because the design gets stuck in a local minimum prioritizing only one metric.

Equality saturation has been shown to be effective in situations like these, where destructive, greedy rewriting leads to a poor global cost. However, we observe empirically that equality saturation struggles to scale with the demands of FHE circuit optimization, often producing considerably worse results than state-of-the-art greedy methods. 
In particular, we find that the equivalence information generated in the e-graph by saturation, does not consistently lead to better solutions during extraction. For MD (which admits efficient optimal extraction), saturation fails to find solutions identified by simpler greedy approaches. Furthermore, the e-graph sizes generated by saturation  make the use of optimal MC extraction methods, like those based on ILP,  impractical or prohibitive to apply.

Instead, we introduce a method we call ``cut tracing'', which uses e-graphs to store the intermediate optimizations found by existing MD and MC reduction algorithms. 
% and the optimal term is extracted?
Cut tracing is able to strike a balance between the tendency of these algorithms to fall into local minima and the inefficiency of equality saturation for problems of this scale. Compared to equality saturation in particular, cut tracing produces e-graphs small enough that we not only gain the ability to use an ILP formulation for MC-optimal extraction, but to add MD constraints to the formulation, allowing us to sweep the design space of MD/MC tradeoffs. 

In this paper, we motivate and discuss cut tracing with e-graphs as a means to optimize Boolean FHE circuits. We use a flow combining the state-of-the-art MD and MC algorithms ESOP balancing and MC-oriented cut rewriting and resubstitution \cite{haner2022esop, testa2020toolbox} as our baseline, showing how cut tracing can augment this flow to minimize the overall HE runtime of a suite of benchmarks.

\section{Background} 

\subsection{XOR-AND graphs}

As mentioned, in the case of a Boolean plaintext space, an FHE circuit is equivalent to a XOR-AND graph (XAG). We define an XAG as a directed acyclic graph with nodes $V$ and edges $E$. An edge $e = (v, v_i) \in E$ corresponds to a fanin $v_i$ for the gate $v$. We define the set of \textit{primary inputs} $PI = \{ v \in V | \forall v_i, (v, v_i) \notin E \} $, and \textit{primary outputs} $PO = \{v \in V | \forall v_o, (v_o, v) \notin E \}$. The nodes in $V-PI$ correspond to the 2-input gates $\wedge$ and $\oplus$. 

Next, we define a function $d(v)$ where $d(v) = 1$ if $v\in V$ is a multiplication, and $d(v) = 0$ otherwise. Then, the \textit{multiplicative depth} (MD) of a node is given recursively by 
$$ MD(v) = d(v) + \max_{c | (v,c) \in E} MD(c). $$
We define the $MD$ of the entire circuit as $MD = \max_{v \in PO} MD(v)$. 
 Similarly, we define the \textit{multiplicative complexity} (MC) of a circuit as  
$$ MC = |\{ v | v \in V, d(v) = 1 \}|$$

\subsection{Cuts}

A cut $C$ of an XAG is a set of \textit{leaves} $L$ and a \textit{root} $v$ such that a) every path from $v$ to a primary input $PI$ passes through at least one leaf, and b) each leaf is contained in at least one such path. We say that a cut $C$ is \textit{$k$-feasible} if it has at most $k$ leaves. 

Cuts are commonly used in logic synthesis algorithms to optimize small pieces of a large logic network at a time. That is, given some cut $C$ of the original network, a $C'$ is found that implements the same Boolean function over the set of leaves $L$, with a lower cost. This method is employed by the MD and MC reducing algorithms we use as a baseline \cite{haner2022esop}\cite{testa2020toolbox}, namely cut \textit{rewriting}, \textit{resubstitution}, and \textit{balancing}. Rewriting and resubstitution are used to reduce MC, while balancing (with ESOP forms) reduces MD. We omit the details of each algorithm here, but provide a high-level overview describing the aspects they share in common, which are most relevant for this work.

\begin{figure*}[htbp]
\begin{subfigure}[b]{0.25\textwidth}
\centering
\begin{tikzpicture}
    \node[state, xshift=0cm, yshift=0cm, minimum size=0.5cm] (a)   {$a$};
    \node[state, xshift=1.3cm, yshift=0cm, minimum size=0.5cm] (b)   {$b$};
    \node[state, xshift=2.8cm, yshift=0cm, minimum size=0.5cm] (cin)   {$c_{in}$};
    \node[state, xshift=0cm, yshift=1.2cm, minimum size=0.5cm] (n1) {$\wedge$};
    \node[state, xshift=1.3cm, yshift=1.2cm, minimum size=0.5cm] (n2) {$\oplus$};
    \node[state, xshift=0.2cm, yshift=2.4cm, minimum size=0.5cm] (n3) {$\oplus$};
    \node[state, xshift=1.5cm, yshift=2.4cm, minimum size=0.5cm] (n4) {$\wedge$};
    \node[state, xshift=2.8cm, yshift=2.4cm, minimum size=0.5cm] (nsum) {$\oplus$};
    \node[state, xshift=0.7cm, yshift=3.6cm, minimum size=0.5cm] (ncout) {$\oplus$};
    \node (sum) at (2.8,4.5) {$sum$};
    \node (cout) at (0.7,4.5) {$c_{out}$};
    
    \draw [->]  (a) edge (n1);
    \draw [->]  (a) edge (n2);
    \draw [->]  (b) edge (n1);
    \draw [->]  (b) edge (n2);
    \draw [->]  (n1) edge (n3);
    \draw [->]  (n2) edge (n3);
    \draw [->]  (n2) edge (n4);
    \draw [->, dashed]  (cin) edge (n4);
    \draw [->]  (n2) edge (nsum);
    \draw [->]  (cin) edge (nsum);
    \draw [->]  (n3) edge (ncout);
    \draw [->]  (n4) edge (ncout);
    \draw [->]  (ncout) edge (cout);
    \draw [->]  (nsum) edge (sum);
    %\draw [->] (x1) edge (n3);
    %\draw [->] (x2) edge (n3);
    %\draw [->] (x2) edge (n4);
    %\draw [->] (x1) edge (n4);
    %\draw [->] (n3) edge (n5);
    %\draw [->] (n4) edge (n5);
    %\draw [->] (n5) edge (f);
\end{tikzpicture}
\caption{Unoptimized full adder XAG, MC = 2}
\label{fig:unopt_fa}
\end{subfigure}
\begin{subfigure}[b]{0.25\textwidth}
\centering
\begin{tikzpicture}
    \node[state, xshift=0cm, yshift=0cm, minimum size=0.5cm] (a)   {$a$};
    \node[state, xshift=1.3cm, yshift=0cm, minimum size=0.5cm] (b)   {$b$};
    \node[state, xshift=2.8cm, yshift=0cm, minimum size=0.5cm] (cin)   {$c_{in}$};

    \node[state, xshift=0.7cm, yshift=1.2cm, minimum size=0.5cm] (n1) {$\oplus$};
    \node[state, xshift=2.0cm, yshift=1.2cm, minimum size=0.5cm] (n2) {$\oplus$};
    \node[state, xshift=0.7cm, yshift=2.4cm, minimum size=0.5cm] (n3) {$\wedge$};
    \node[state, xshift=2.6cm, yshift=2.4cm, minimum size=0.5cm] (n4) {$\oplus$};
    \node[state, xshift=0cm, yshift=3.6cm, minimum size=0.5cm] (n5) {$\oplus$};    
    \node (sum) at (2.6,4.5) {$sum$};
    \node (cout) at (0,4.5) {$c_{out}$};
    
    \draw [->]  (a) edge (n1);
    \draw [->]  (b) edge (n1);
    \draw [->]  (b) edge (n2);
    \draw [->]  (cin) edge (n2);
    \draw [->]  (n1) edge (n3);
    \draw [->, dashed]  (n2) edge (n3);
    \draw [->]  (n1) edge (n4);
    \draw [->]  (cin) edge (n4);
    \draw [->]  (n3) edge (n5);
    \draw [->]  (a) edge (n5);
    \draw [->]  (n4) edge (sum);
    \draw [->]  (n5) edge (cout);
    
\end{tikzpicture}
\caption{MC=1 optimal full adder}
\label{fig:opt_fa}
\end{subfigure}
\begin{subfigure}[b]{0.25\textwidth}
 \centering
 \includegraphics[width=\textwidth,height=0.8\textwidth]{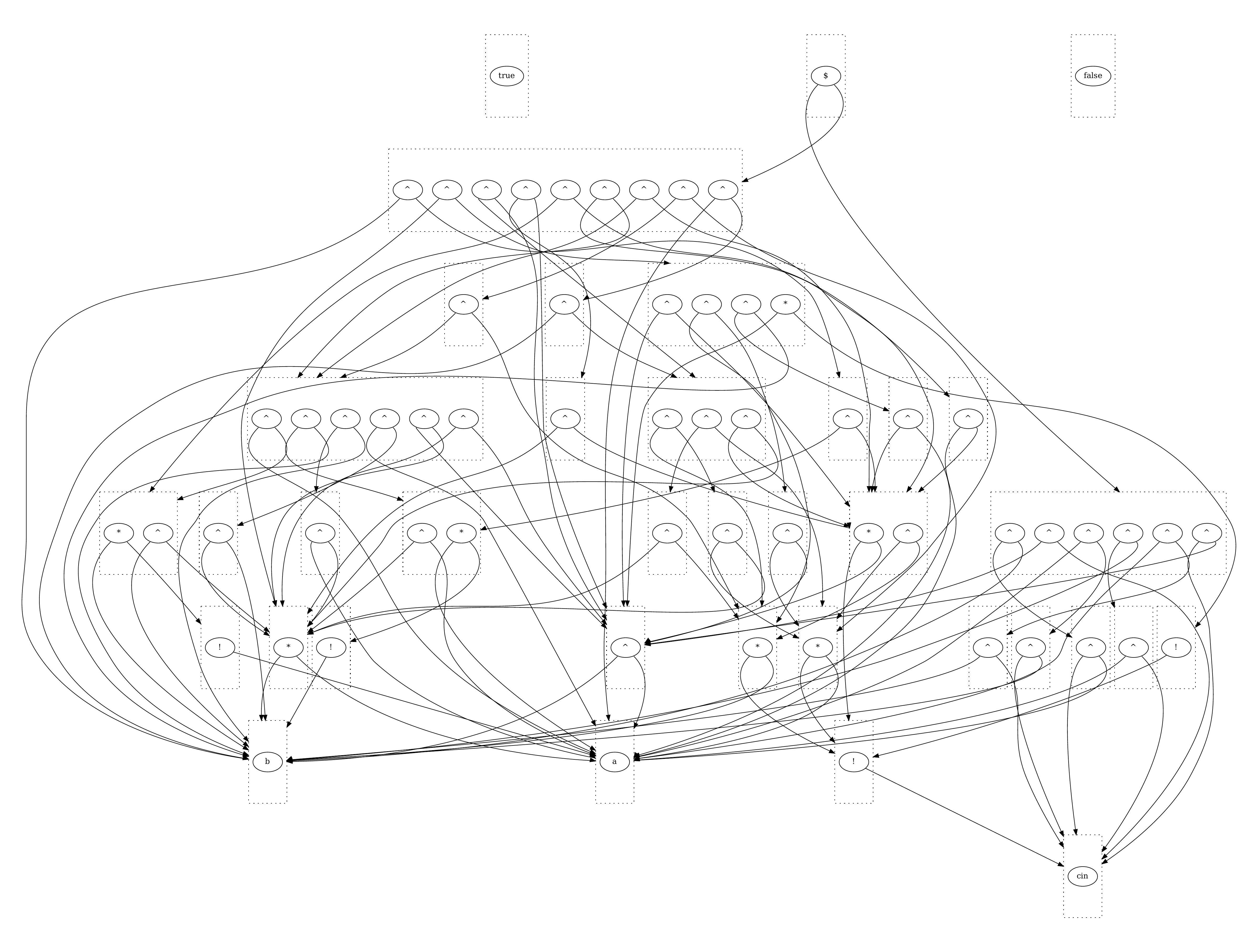}
 \caption{Full adder circuit in (a) after 2 iterations of eqsat. MC reported by ILP = 2}
\label{fig:fa_egg}
\end{subfigure} 
\caption{XAGs representing a full adder circuit, along with the e-graph after 2 iterations in \textit{egg} using Boolean laws as rewrite rules. Dashed edges represent complement i.e. a NOT gate, which is considered to be free in terms of FHE as it can be implemented with XOR. }
\end{figure*}
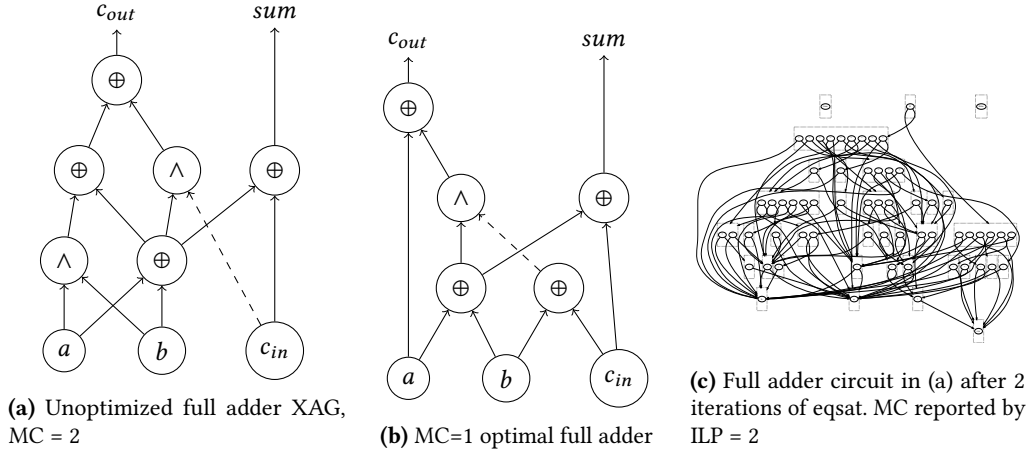 

\subsection{Rewriting, resubstitution, balancing}

At a high-level, each of these algorithms performs peephole optimization through cuts on a logic network. The optimization is guided through the compuation of a \textit{gain}: the effect of the optimized cut on the cost of the network, estimated locally. Algorithm \ref{alg:cut_opt} shows the structure shared by each. The behavior of the functions $OPT$, $CUTS$, and $GAIN$ are dependent on the algorithm and the mode of optimization. For MC-oriented cut rewriting, for example, the optimization function $OPT$ looks up known the Boolean function corresponding to a 6-feasible cut in a database of known MC-optimal XAG implementations for 6-input functions \cite{testa2020toolbox}. In ESOP balancing, $OPT$ expresses the cut in an exclusive-sum-of-products form, which can be used to reduce the number of AND levels ($MD$) of the Boolean function \cite{haner2022esop}. For $CUTS$, balancing and rewriting both make use of a procedure called \textit{cut enumeration}, which recursively enumerates $k$-feasible cuts given some fixed $k$. Finally, in the case of balancing, $GAIN$ is given by the difference in $MD$ of the cuts, while resubstitution and rewriting formulate gain as the the saving in $MC$ when the cut is replaced in the graph.

\begin{algorithm}
\caption{Structure of cut-based optimization}
\begin{algorithmic}[1]
\Require Logic network $N=(V,E)$, optimization function $OPT(C)$, cut identifying function $CUTS(n)$, gain function $GAIN(C)$
\For{node $v \in V$}
    \State $(G_{best}, C_{best}) \gets (0, \text{None})$
    \For{cut $C \in CUTS(v)$}
        \State $C_{opt} \gets OPT(C)$
        \State $G_{opt} \gets GAIN(C_{opt})$
        \If{$G_{opt} > G_{best}$}
            \State $(G_{best}, C_{best}) \gets (G_{opt}, C_{opt})$
        \EndIf
    \EndFor
    \If{$G_{best} > 0$}
        \State replace $C_{best}$ at $v$
    \EndIf
\EndFor
\end{algorithmic}
\label{alg:cut_opt}
\end{algorithm}

\section{Cut Tracing}

In this section, we motivate and describe our approach to optimizing FHE circuits with e-graphs, called cut tracing.

\subsection{Motivation}

A key weakness of the balancing, rewriting and resubstitution algorithms for MD and MC discussed above is that, among the cuts evaluated, the replacement cut with the highest gain is always selected, and this choice is never revisited. This is especially problematic when trying to combine the two flows to achieve an MD/MC balanced circuit: the design is likely to already be stuck in a local minimum. 

Equality saturation offers a compelling solution to this problem. Instead of optimizing directly on the cuts of the network, destroying opportunities for potentially superior designs, we use an e-graph combined with rewrite rules to collect a rich dataset of equivalent representations. We expect that this e-graph will eventually become saturated with the optimal design, which we can then extract out.

However, we observe that for FHE circuits, in practice, equality saturation fails to keep up with the efficiency of ESOP balancing and MC optimization. We demonstrate with a small example. Figure \ref{fig:unopt_fa} depicts an XAG of a full adder circuit, with a multiplicative complexity of 2. For small circuits such as these, the exact optimal MC is known, and in this case the optimal version requires only 1 AND gate (Figure \ref{fig:opt_fa}.) We attempt to use equality saturation to find this design. Inspired by previous work on equality saturation for a different logic synthesis problem\cite{chen2024esyn}, we use a similar set of Boolean laws over AND, XOR, and NOT as rewrite rules, such as distributivity, redundancy, consensus, etc.

Figure \ref{fig:fa_egg} shows that in just 2 iterations of saturation using the \texttt{egg} equality saturation framework~\cite{willsey2021egg}, the size of the e-graph explodes to contain 37 e-classes and 63 e-nodes (up from 10 nodes in the original graph), and still has not found the optimal MC. By 5 iterations, with 99 e-classes and 394 e-nodes, ILP finally reports an MC of 1. By contrast, cut rewriting can identify this design in a single step: since this whole circuit could be considered a 6-feasible cut, it appears in the database of MC-optimal 6-input functions. Compounded over an entire circuit, the efficiency loss of having to find optimizations through repeated restructuring rewrites is significant.

One solution to these problems is to use coarser rewrite rules which similarly replace ``cuts'' of the e-graph at a time. While this can solve the inefficiency of the simple rules, we saw it introduce a new problem: now the e-graph needs to be able to match against the left-hand side. Since there are many different equivalent ways to structure a given XAG (AND/XOR are commutative and associative), the results depend on equality saturation's ability to expose the structure of the left-hand-side in the e-graph. In addition, this does not solve the issue that equality saturation performs an undirected search of rewriting rules. To apply 1 rule which is used in the final solution, it may apply a multitude of others which add unviable e-classes and e-nodes. At the scale of hundreds or thousands of nodes present in typical FHE circuits, this becomes a problem for the runtime of exact-optimal DAG cost extraction methods like ILP, forcing the use of heuristics that compromise the ability to achieve optimal MC.

The bottleneck we observe in equality saturation is in the need to restructure the e-graph to expose rewriting opportunities. On the other hand, techniques like MC optimization and ESOP balancing completely circumvent this issue. They do not use structural rewriting at all, but rather identify a replacement cut implementing the same Boolean function, for each cut considered in the network. But the relationship between a cut and its replacement is an equivalence that can also be recorded in an e-graph. In doing so, the explosion of the e-graph size through restructuring is avoided, while compactly representing a rich set of local design choices that can be passed to extraction to minimize the global cost. Based on this insight, we propose cut tracing: recording, or ``tracing'' the equivalence collected by a cut-based optimization algorithm. 

\subsection{Description}

\begin{figure}
\includegraphics[width=0.45\textwidth,height=0.45\textwidth]{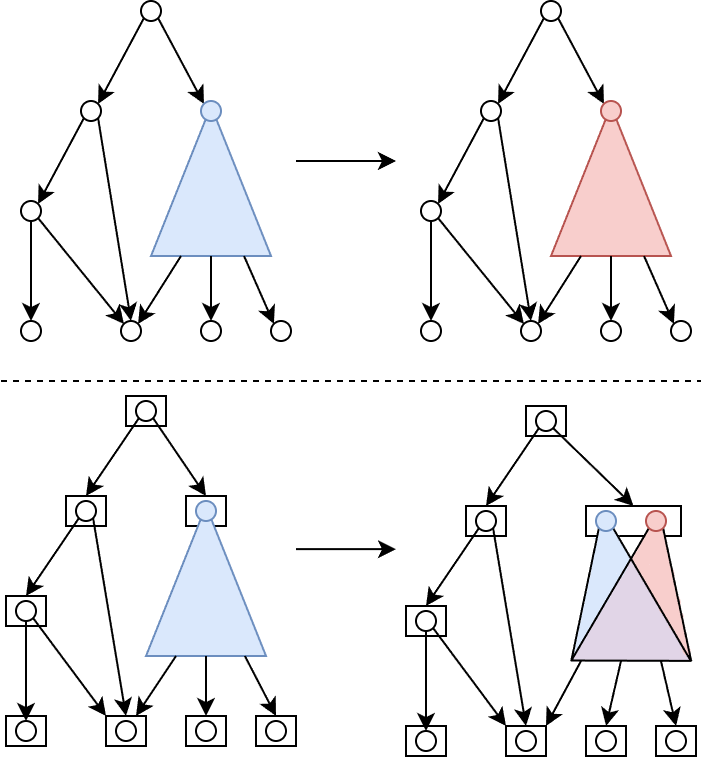}
\caption{A visualization of cut tracing. The top shows a cut of a logic network being replaced, while the bottom shows the operation being traced in the e-graph by adding the new cut root to the same e-class. The purple region denotes sharing between the two cuts.}
\label{fig:cut_tracing_viz}
\end{figure}

Cut tracing augments an existing algorithm that manipulates a circuit by optimizing cuts. The behavior of cut tracing is simple: an e-graph is first initialized to hold the input circuit. The algorithm to which cut tracing is attached proceeds as normal, with the caveat that whenever a replacement cut is identified, it is also added to the e-graph, and the roots of the original and replacement cuts are placed in the same e-class. This process happens alongside the replacement in the original network carried out by the algorithm normally. Figure \ref{fig:cut_tracing_viz} visualizes the process. Note that the original and replacement cut may share some structure, shown in the purple region, which the e-graph automatically handles through hashconsing.

When the algorithm completes, like in equality saturation, the optimal term can be extracted from the e-graph with the traced cuts, according to the desired cost function. Crucially, however, tracing also works alongside the \textit{composition} of multiple algorithms, which may each optimize different parameters (like MD and MC.) By passing the same e-graph along each optimization step, information is never destroyed, only added. We use this property in the context of FHE circuits to apply both MC optimization and ESOP balancing in sequence, while retaining enough information to allow decisions made by either that are detrimental to overall HE cost to be ``undone'' through extraction.

As mentioned, cut tracing does not interfere with the optimizations that the algorithm would normally produce. Consequently, notwithstanding extraction limitations, its results must be at least as good as the network produced by the original sequence of optimizations, since this form must exist in the e-graph. However, this heavily constrains its ability to expose better designs. Cut tracing cannot explore the effect of applying optimizations to a cut that was never enumerated in the source network. We show that in spite of these limitations, with just the information collected from tracing the standard behavior of these algorithms, the e-graph can frequently recover a more balanced design.

While we show its efficacy for FHE circuits, we note that cut tracing could be applied to any logic synthesis algorithm that is based on cuts, of which there are many, since optimization strategies which scale to entire networks are limited. 

\section{Discussion \& Evaluation}

In this section, we present our evaluation of the effect of attaching cut tracing to a flow for FHE circuit optimization. We first describe the flow, then compare its effect on the homomorphic evaluation runtime of a suite of FHE benchmarks to the baseline.  

\subsection{Cut Tracing Flow}

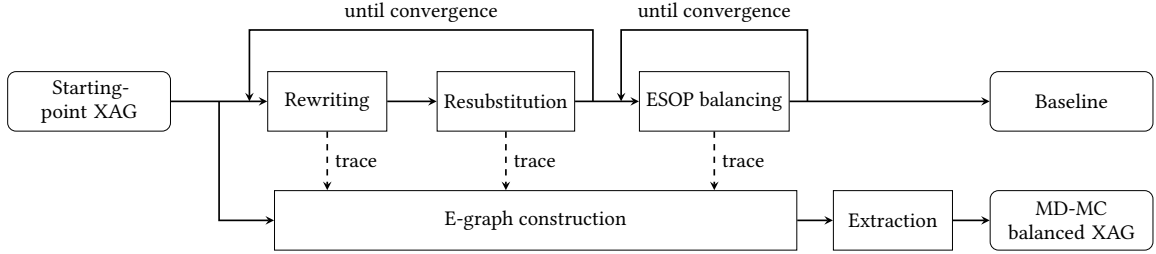
\begin{figure*}[htbp]
    \centering
    \resizebox{0.9\textwidth}{!}{
        \usetikzlibrary{positioning,calc}

\tikzstyle{startstop} = [rectangle, rounded corners, minimum width=1.5cm, minimum height=1cm, text centered, text width=2.5cm, draw=black]
\tikzstyle{process} = [rectangle, minimum width=2cm, minimum height=1cm, text centered, align=center, draw=black]
\tikzstyle{arrow} = [thick,->,>=stealth]

\begin{tikzpicture}[node distance=2cm and 1.7cm]
    \node (start) [startstop] {Starting-point XAG};
    \node (rewrite) [process, right of=start, xshift=2cm] {Rewriting};
    \node (resub) [process, right of=rewrite, xshift=1cm] {Resubstitution};
    \node (rebalance) [process, right of=resub, xshift=1.5cm] {ESOP balancing};
    \node (baseline) [startstop, right of=rebalance, xshift=4cm] {Baseline};
    \path let
        \p1 = (rewrite.west),
        \p2 = (rebalance.east)
    in node (egraph) [process, minimum width=\x2-\x1, xshift=0.5cm, below of=resub, anchor=center] 
        {E-graph construction};
    \node (extract) [process, right of=egraph, xshift=4cm] {Extraction};
    \node (optimal) [startstop, below of=baseline] {MD-MC balanced XAG};
    
    \draw [arrow] (start) -- (rewrite);
    \draw [arrow] (rewrite) -- (resub);
    \draw [arrow] (resub) -- (rebalance);
    \draw [arrow] (rebalance) -- (baseline);
    \draw [arrow] (egraph)  -- (extract);
    \draw [arrow] (extract) -- (optimal);
    \path (start) -- (rewrite) coordinate[pos=0.5] (midpoint);
    \draw [arrow] (midpoint) |- (egraph);
    \draw [arrow, dashed] (rewrite.south) -- node[right] {trace} (rewrite.south |- egraph.north);
    \draw [arrow, dashed] (resub.south) -- node[right] {trace} (resub.south   |- egraph.north);
    \draw [arrow, dashed] (rebalance.south) -- node[right] {trace} (rebalance.south |- egraph.north);
    \draw[arrow]
        ([xshift=0.3cm]resub.east)
        -- ++(0,1.2)
        -- node[above]{until convergence} ([xshift=-0.3cm, yshift=1.2cm]rewrite.west)
        -- ([xshift=-0.3cm]rewrite.west);
    \draw[arrow]
        ([xshift=0.3cm]rebalance.east)
        -- ++(0,1.2)
        -- node[above]{until convergence} ([xshift=-0.3cm, yshift=1.2cm]rebalance.west)
        -- ([xshift=-0.3cm]rebalance.west);
\end{tikzpicture}
    }
\caption{MC + MD optimization flow incorporating cut tracing.}
\label{fig:flow}
\end{figure*}

Figure \ref{fig:flow} shows our proposed flow. Cut rewriting and resubstitution are first applied to the circuit in sequence, reducing the MC until no further optimizations are possible (convergence is reached.) Next, ESOP balancing is applied to reduce the MD, also until convergence. The result serves as our baseline. We note that we also intended to test cut tracing with the opposite ordering, ESOP balancing first and MC optimization next, since we observed the choice of the first optimization to impact the results. However, due to issues in our implementation, we do not yet have results for cut tracing on this flow. As a result, we compare cut tracing to the best result achieved by either ordering in the baseline.

In the flow, cut tracing records cuts identified by each algorithm in an e-graph, which is kept between applications of algorithms. One choice that cut tracing does not define is which cuts to record; recording every candidate cut regardless of whether it improves the cost may generate considerable noise of unviable candidates. Based on empirical observations, for cut rewriting \& resubstution, we choose to record only the best cut (maximizing the local gain) identified for each node. For ESOP balancing, however, we record all cuts, even those that do not improve the depth locally, as we found that cut tracing can take advantage of these options. After the flow completes and the e-graph is filled with traced candidates, it is extracted to yield the optimized circuit.

Our extraction method targets \textit{HE cost}, an approximation of the runtime of FHE circuits calculated from MD and MC as $MD^2 \times MC$, based on analyses of the cost of FHE runtime in the literature \cite{costache2015cost}. Algorithm \ref{alg:he_extraction} describes the extraction method we use to minimize HE cost. First, a greedy bottom-up extraction method which accurately evaluates shared MC cost of visited terms, but does not account for globally optimal MC, is used. The algorithm  also prioritizes MD over MC, meaning it returns an MD optimal circuit but uses MC to break ties. $md\_global\_{greedy}\_dag$ modifies the implementation available in \cite{extraction_gym} to incorporate $MD$ prioritization. 

Next, we incorporate an extension to the standard e-graph ILP extraction formulation with constraints to ensure the solution does not not exceed a bound on MD ($depth\_bounded\_ilp$). This provides an MC-optimal solution with respect to a given MD bound.  Using this, the algorithm sweeps a fixed range of possible MD values, continuously relaxing the bound on MD to locate the overall best solution in terms of HE cost. We note that the addition of depth constraints adds non-insignificant solver runtime to the ILP formulation, but the small e-graph sizes produced by tracing means that this extraction technique remains viable.

\begin{algorithm}
\caption{Extraction for HE cost}
\begin{algorithmic}[1]
\Require E-graph \(E\), number of ILP iterations \(k\)
\Ensure Extracted XAG \(N\)
\State \( (N_{best}, MD_{best}, MC_{best}) \gets md\_global\_greedy\_dag(E) \)
\State \( C_{best} \gets MD_{best}^2 \times MC_{best} \)
\For{$i = 0 \to k$}
    \State \( (N, MC) \gets depth\_bounded\_ilp(E, MD_{best}+i) \)
    \State $C_i \gets (MD_{best}+i)^2 \times MC$
    \If{$C_i \leq C_{best}$} 
        \State $(N_{best}, C_{best}) \gets (N,C_i)$
    \EndIf
\EndFor
\State \Return $N_{best}$
\end{algorithmic}
\label{alg:he_extraction}
\end{algorithm}

\begin{table*}[!t]
\begin{center}
\begin{tabular}{l | r r r | r r r | r r r | r r }
\hline
& \multicolumn{3}{c}{\textbf{Baseline}} & \multicolumn{3}{|c}{\textbf{Baseline (MD first)}} & \multicolumn{3}{|c}{\textbf{Cut Tracing}} & &  \\ \hline
\textbf{Name} & \textbf{\textit{MD}} & \textbf{\textit{MC}} & \textbf{\textit{Eval. [s]}} & \textbf{\textit{MD}} & \textbf{\textit{MC}} & \textbf{\textit{Eval. [s]}} & \textbf{\textit{MD}} & \textbf{\textit{MC}} & \textbf{\textit{Eval. [s]}} & \makecell{\textbf{\textit{Speedup}} \\ \textbf{\textit{(baseline)}}} & \makecell{\textbf{\textit{Speedup}} \\ \textbf{\textit{(best order)}}}\\\hline
bar & 10 & 1107 & 100.5 & 10 & 1107 & 101.7 & 10 & 1115 & 106.5 & 0.94 & 0.94 \\ \hline 
bsort & 45 & 390 & 723.2 & 43 & 498 & 649.0 & \blue{43} & \blue{394} & \blue{535.6} & \blue{1.35} & 1.21 \\ \hline 
cardio & 8 & 93 & 6.6 & 8 & 82 & 6.0 & 8 & 80 & 5.9 & 1.10 & 1.01 \\ \hline 
cavlc & 11 & 623 & 69.2 & 11 & 474 & 54.8 & 11 & 638 & 72.4 & 0.96 & 0.76 \\ \hline 
ctrl & 4 & 89 & 2.5 & 5 & 50 & 1.5 & 4 & 82 & 2.4 & 1.06 & 0.62 \\ \hline 
dec & 3 & 292 & 4.7 & 3 & 292 & 4.7 & 3 & 292 & 4.7 & 0.99 & 0.99 \\ \hline 
dsort & 8 & 594 & 37.9 & 7 & 540 & 28.9 & 8 & 594 & 38.8 & 0.98 & 0.75 \\ \hline 
hd01 & 5 & 107 & 2.8 & 5 & 85 & 2.3 & 5 & 83 & 2.2 & 1.29 & 1.04 \\ \hline 
hd02 & 7 & 84 & 4.5 & 6 & 76 & 3.67 & 6 & 67 & 3.2 & 1.40 & 1.15 \\ \hline 
hd03 & 5 & 34 & 0.9 & 5 & 27 & 0.8 & 5 & 26 & 0.7 & 1.29 & 1.03 \\ \hline 
hd04 & 8 & 58 & 4.0 & 8 & 50 & 3.6 & 8 & 41 & 3.0 & 1.33 & 1.21 \\ \hline 
hd05 & 7 & 105 & 4.7 & 6 & 138 & 6.9 & 7 & 105 & 4.8 & 0.97 & 0.97 \\ \hline 
hd06 & 7 & 105 & 4.7 & 6 & 138 & 6.8 & 7 & 105 & 4.8 & 0.98 & 0.98 \\ \hline 
hd07 & 4 & 11 & 0.3 & 5 & 17 & 0.43 & 4 & 11 & 0.3 & 0.97 & 0.97 \\ \hline 
hd08 & 5 & 12 & 0.3 & 6 & 18 & 0.84 & 5 & 12 & 0.3 & 0.97 & 0.97 \\ \hline 
hd09 & 10 & 135 & 11.8 & 11 & 115 & 12.4 & 10 & 95 & 8.4 & 1.40 & 1.40 \\ \hline 
hd10 & 5 & 34 & 0.9 & 5 & 32 & 1.0 & 5 & 32 & 0.9 & 1.02 & 1.02 \\ \hline 
hd11 & 14 & 384 & 76.1 & 14 & 345 & 74.3 & 14 & 341 & 69.7 & 1.09 & 1.07 \\ \hline 
hd12 & 14 & 70 & 13.8 & 14 & 70 & 13.6 & 14 & 70 & 14.1 & 0.98 & 0.96 \\ \hline 
i2c & 9 & 1112 & 81.2 & 11 & 1039 & 118.1 & 9 & 1082 & 81.1 & 1.00 & 1.00 \\ \hline 
int2float & 9 & 218 & 15.6 & 9 & 180 & 13.3 & 9 & 223 & 16.6 & 0.94 & 0.80 \\ \hline 
isort & 45 & 390 & 722.1 & 43 & 498 & 651.4 & 43 & 394 & 539.8 & 1.34 & 1.21 \\ \hline 
msort & 45 & 390 & 722.5 & 43 & 498 & 688.6 & 43 & 394 & 539.0 & 1.34 & 1.28 \\ \hline 
osort & 25 & 338 & 154.7 & 25 & 338 & 156.8 & 25 & 338 & 163.7 & 0.94 & 0.94 \\ \hline 
router & 12 & 213 & 27.6 & 12 & 224 & 29.4 & 12 & 164 & 21.6 & 1.28 & 1.28 \\ \hline
\multicolumn{10}{l}{\textit{Total (geomean)}} & \textbf{1.10} & \textbf{1.00} 
\end{tabular}
\caption{MD, MC and homomorphic evaluation time of benchmark suite, for baseline in both MD-first and MC-first orders, and MC-first with cut tracing enabled.}
\label{tab:evalresults}
\end{center}
\end{table*}

\subsection{Experimental Setup}

All experiments were conducted on an M1 MacBook Pro with 16GB RAM.

\textbf{Implementation.} We implement cut tracing atop the logic synthesis library \texttt{mockturtle}, which provides the implementation of cut rewriting, resubstitution, and ESOP balancing \cite{libraries}. We added hooks to these algorithms to record the equivalences needed to reconstruct an e-graph in a trace file. We implement the replaying of the trace and the extraction of Algorithm \ref{alg:he_extraction} using the \texttt{egg} library \cite{willsey2021egg}. 

%To test equality saturation, we also use \texttt{egg}.

To evaluate the FHE circuit runtime, we use the \texttt{HELib} FHE library \cite{shai2020helib}. Security parameters are chosen by \texttt{HELib} according to a Boolean plaintext space, and the depth of the input circuit, with the security level $\lambda = 128$. The circuit evaluator traverses the input circuit in a bottom-up fashion, serially executing each AND or XOR node in the network. Orthogonal FHE optimizations such as batching are not utilized, and the evaluator conservatively relinearizes the ciphertexts after each AND node.

\textbf{Benchmarks.} We source a set of Boolean FHE circuit benchmarks from \cite{lee2020lobster}, which represent a mix of standard combinational logic circuits in XAG form, as well as applications of interest to FHE such as sorting and medical diagnosis. 

\textbf{Baseline.} For resubstitution and rewriting, we use the same parameters as \cite{testa2020toolbox}. For ESOP balancing, we reuse the same cut enumeration parameters (6-input cuts with 12 cuts per node.) ESOP balancing is only applied on the critical path of the circuit.

\textbf{E-Graph Setup.} For HE cost extraction, we fix the number of iterations $k$ to 2, observing that the best HE cost design is usually found close to the optimal depth. We use a timeout of 10 minutes for the ILP solver.

%For equality saturation, we use the set of rules \ref{tab:boolrules} discussed earlier. For space, we do not test the HE runtime of the circuits produced by equality saturation, only showing the relationship between MD and MC. 

\subsection{Results} 

Table \ref{tab:evalresults} displays our results. We record the MD, MC, and homomorphic evaluation runtime of each benchmark. We see that cut tracing is frequently able to find improvements over the flow it is applied to, usually in the form of MC reduction. We highlight \texttt{bsort} in particular (in \blue{blue}), where cut tracing was not only able to recover a lower MD, but also barely trade off MC in the process (4 AND gates). In fact, the e-graph for \texttt{bsort} contained an MD of 42, but as the ILP solver timed out with this as the depth constraint, HE cost extraction was able to identify this solution with a relaxed bound of 43.

Other benchmarks end up with a worse overall solution in both MD and MC, but as mentioned, this should not be possible with the model of cut tracing, where the design reported by the standard optimization flow should exist in the e-graph. We believe this issue is due to the way the default ILP formulation in \texttt{egg} handles the acylicity constraint, which is to ban the use of any e-nodes detected as being part of a cycle. Accordingly, every benchmark which did not improve the baseline had cycles, and those that did had few or none. Future work will integrate existing exact optimal extraction algorithms with more sophisticated handling for cycles, such as enforcing topological sorting in ILP constraints. 

Comparing to the ESOP first ordering, we see firstly that, as expected, the ordering of MD and MC flows heavily influences the path the design takes. One order is not always the best. Secondly, we can see that in several cases (\*\texttt{sort}, \texttt{router}, \texttt{hd04}) where ESOP first is the better order, cut tracing is still able to beat out this design, making it the overall best. However, it is still heavily constrained to the design choices recorded in the original flow, as it is not able to find a superior design identified by ESOP first in other cases.

\section{Conclusion}

We presented cut tracing, a logic optimization technique using e-graphs that trades off the wider design space exploration possible with equality saturation for greater scalability. We applied cut tracing towards the application of Boolean FHE circuit synthesis, creating a flow that explicitly optimizes for overall HE runtime, rather than the individual metrics of MC and MD. Cut tracing extends existing logic synthesis algorithms, rather than replacing them, and we showed that adding it to a sequence of existing flows for MD and MC optimization yields a 40\% best-case and 10\% average-case speedup over the baseline. Due to implementation issues preventing us from testing cut tracing on other flows, it is only 20\% best-case and tied over our best-known results. We plan to fix these issues and continue investigating cut tracing, specifically in terms of improving extraction and in ways the e-graph can be incorporated during the design flow, not just orthogonally. 

\bibliographystyle{ACM-Reference-Format}
\bibliography{biblio}

\end{document}